\documentclass[11pt]{article}

\usepackage{amsfonts,amsmath,amssymb,graphics}

\title{\bf Mean Reversion Pays, but Costs\footnote{This is a longer version of an article published in RISK(24)2:84--89 (Feb.~2011).}}

\newcommand{\C}{\mathbb{C}}

\newcommand{\LL}{\mathcal{L}}

\newcommand{\ex}{\mathbf{E}}

\newcommand{\util}{\mathcal{U}}

\newcommand{\notthis}[1]{}

\newcommand{\half}{\frac{1}{2}}

\newcommand{\cdl}{\,|\,}

\newcommand{\shalf}{{\textstyle\frac{1}{2}}}

\newcommand{\pderiv}[2]{\frac{\partial{#1}}{\partial{#2}}}
\newcommand{\deriv}[2]{\frac{d{#1}}{d{#2}}}
\newcommand{\dderiv}[2]{\frac{d^2{#1}}{d{#2}^2}}
\newcommand{\ddderiv}[2]{\frac{d^3{#1}}{d{#2}^3}}

\newcommand{\pdderiv}[2]{\frac{\partial^2{#1}}{\partial{#2}^2}}

\newcommand{\money}{G}
\newcommand{\hd}{\hat{h}_d}
\newcommand{\invar}{W}
\newcommand{\cutstart}{}
\newcommand{\cutend}{}
\newcommand{\tcmbuy}{\varepsilon_+}
\newcommand{\tcmsell}{\varepsilon_-}

\begin{document}
%\AHLtitle
\maketitle

\begin{abstract}

A mean-reverting financial instrument is optimally traded by buying it when it is sufficiently below the estimated `mean level' and selling it when it is above. In the presence of linear transaction costs, a large amount of value is paid away crossing bid-offers unless one devises a `buffer' through which the price must move before a trade is done. In this paper, Richard Martin and Torsten Sch\"oneborn derive the optimal strategy and conclude that for low costs the buffer width is  proportional to the cube root of the transaction cost, determining the proportionality constant explicitly.

\end{abstract}

%%%%%%%%%%%%%%%%%%%%%%%%%%%%%%%%%%%%%%%%%%%%%%%%%%%%%%%%%%%%%%%%%%%%%%%%%%%%%%%%%%%%%%%%%%%%%%%%%%%%%%%%%%%%%%%%%%%%%%%%%%%%%%%%%%%%%%%%%%%%%%%%%%%%
%%%%%%%%%%%%%%%%%%%%%%%%%%%%%%%%%%%%%%%%%%%%%%%%%%%%%%%%%%%%%%%%%%%%%%%%%%%%%%%%%%%%%%%%%%%%%%%%%%%%%%%%%%%%%%%%%%%%%%%%%%%%%%%%%%%%%%%%%%%%%%%%%%%%

\section*{Introduction}

A difficult problem in trading algorithm design is linear transaction costs.
This problem is quite distinct from, and much less analytically tractable than, quadratic costs \cite{Garleanu09}, and unless very large positions are being traded it is the major source of slippage.

\begin{figure}[h!]
\centerline{\input{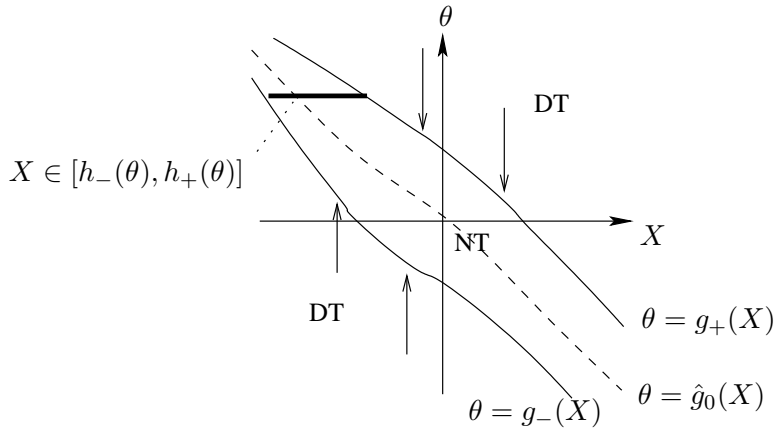}}
\caption{\small Sketch of DT-NT-DT form of optimal trading strategy. In the NT zone, no trade is done. Outside, a trade is done to the edge of the boundary, as implied by the arrows. The dashed line is the optimal $\theta$ in the costfree case.}
\label{fig:dtntdt}
\end{figure}

Traditionally the problem has been considered in the context of trading a stock in a portfolio consisting of stock and a risk-free bond (the so-called Merton problem). Loosely, in the Merton problem, to achieve optimal utility one needs to maintain a fixed proportion of value in the stock, which necessitates continuous trading. Without attention to linear costs, one would incur in a time step $dt$ a cost of order $\sqrt{dt}$, so ``any [literal] attempt to apply Merton's strategy in the presence of transaction costs would result in immediate penury'' (in Davis \& Norman's words \cite{Davis90}). A threshold is therefore constructed through which the price has to move before rebalancing is done. The idea translates into the trading of an arbitrary asset as follows. If its present value $X$ is plotted against the current position $\theta$, the $(X,\theta)$ plane is divided into three zones (Figure \ref{fig:dtntdt}): a no-trade zone (NT) in the middle, and on each side, discrete-trade zones (DT) in which the optimal strategy is to trade directly to the boundary. A strategy of this form is said to be of DT-NT-DT type for obvious reasons.

The Merton problem has been tackled by several authors.
Davis \& Norman \cite{Davis90} examine it probabilistically and derive the optimal buffer shape.
%\footnote{Incidentally they contend that the derivation of the optimal solution required techniques not available to previous authors \cite{Magill76} examining the problem in the 1970s, though we tentatively dispute this, as the techniques used here are completely 19th-century.}
Shreve \& Soner \cite{Shreve94} examine the same problem using viscosity solutions of PDEs to derive their results. A variant has been tackled by Whalley \& Wilmott \cite{Whalley97} and Zakamouline \cite{Zakamouline06} in the hedging of options under linear transaction costs. Intriguingly, both sources of problem produce the same conclusion to the extent that the width of the NT region is, for small transaction costs, proportional to the cube root of the cost of trading one lot of the underlying asset.

Our encounter with the linear transaction cost problem has been through mean-reversion and the trading of putatively stationary combinations of instruments, which has received impetus in recent years through, for example, the theory of cointegration (see e.g.\ \cite{Hamilton94} for a discussion). The existence of such stationary combinations is a form of potentially exploitable market inefficiency and is discussed by Boguslavsky \& Boguslavskaya \cite{Boguslavsky04} who do not treat transaction costs at all. Whereas in the context of the Merton problem, one can circumvent the transaction cost problem simply by rebalancing the portfolio only infrequently \cite{Rogers01}, in mean-reverting strategies one is \emph{forced} to trade frequently, as that is the only way of making money\footnote{In the absence of `carry'.}. So the transaction-cost problem has to be fixed rather than side-stepped.

In this paper we make a number of innovations on the classical case considered by Davis \& Norman. Rather than dealing with `cash' instruments, we assume them to be `synthetic', for example swaps or futures. As an example, one might consider a mean-reverting strategy in long-dated vs short-dated government bonds. We set this up by taking a combination of futures contracts, maintaining a small margin account and permitting considerable leverage, rather than trading the underlying securities. In writing down the dynamics of this `combined instrument' (\ref{eq:assetdyn}) we have to make some changes, for several reasons: there will be no risk-free rate in the drift, we wish to introduce mean reversion in the drift, and the volatility will be permitted to be level-dependent\footnote{Spot-dependent, but not time-dependent, local volatility.}. Although we could give an explicit derivation for one particular case such as Ornstein-Uhlenbeck (OU), which is a well-understood stochastic process, it turns out to be scarcely more analytically tractable than the general case, so we deal with that and then give the OU result as an immediate consequence.
We also require a definition of utility that is more suited to trading, being in essence the total discounted expected return reduced by terms related to the integrated variation of P\&L; in other words we are more interested in utility of \emph{incremental P\&L} than of wealth. The same approach is, incidentally, adopted by Brandt et al.\ in a different context \cite{Brandt09}. For reasons that we explain later, we use constant absolute risk aversion.
Finally we note that our model requires different boundary conditions: obviously there cannot be a `no-shorting' condition, particularly as to trade one unit of the combined instrument will probably require being long one future and short another. 

The trading of this mean-reverting synthetic asset gives rise to a transaction cost problem, for which we derive the optimal Markovian solution of DT-NT-DT form\footnote{Incidentally we do not prove here that there is no better strategy of some other type: for example a non-Markovian one in which the optimal trade did not depend simply on the market value and the current position. The reader is therefore asked to take this on trust. Incidentally a non-Markovian solution to the problem might well be rather difficult to implement.}. Remarkably, the expression for the optimal DT-NT boundary is in reasonably closed form, being described by the solution of a pair of coupled nonlinear equations (\ref{eq:obp1+},\ref{eq:obp1-}).

Next we perform a perturbative analysis for small transaction costs, and give a simple explicit expression for the approximate width of the NT zone, finding it to be proportional to the cube root of the transaction cost. This thereby corroborates the previously-mentioned work in a broader setting.

Finally we provide some numerical examples, comparing the optimal boundary with the perturbation approximation. Incidentally the optimal boundary may also be obtained numerically by the method of dynamic programming, thereby giving an independent verification of the optimal boundary equations.

The techniques we use are very straightforward, essentially reducing the problem to a linear ordinary differential equation with boundary conditions, whose solution we then manipulate; however the algebraic hurdles, particularly in the perturbation theory, are substantial and have been compressed here.

\notthis{
The `road-map' is:
\begin{itemize}
\item
Set up the problem as an infinite-horizon utility maximisation problem, and derive the expression for the value function in the absence of transaction costs: this is straightforward and essentially is done to introduce all the notation.
\item
Assuming a solution of DT-NT-DT form with a given but arbitrary shape of boundary, solve for the value function.
\item
Solve for the optimal boundary by differentiating the value function w.r.t.\ the boundary position and setting that derivative to zero.
\item
Develop in a power series in the transaction cost parameter, $\varepsilon$, thereby arriving at the ``cube root law'' (NT zone width $\propto \varepsilon^{1/3}$).
\item
Plot the boundaries for some examples, and compare with numerical solution by dynamic programming.
\end{itemize}

One of the cases we consider is the Ornstein-Uhlenbeck (OU) process. Despite its apparent simplicity it is difficult to analyse in the presence of linear transaction costs: it requires the use of special functions, so little extra work is needed to solve the problem in greater generality and then do OU as a particular case.
} %end notthis

\section*{Notational preliminaries; Costfree solution}

Let $X_t$ be the present value of the aforementioned `combined instrument' that we believe to be mean-reverting. For example, $X_t$ might be a long-short combination of long-dated government bonds, or of stocks, or of commodity futures. The P\&L comes from differences $X_{t+dt}-X_t$, and given that $X_t$ is the PV of a synthetic instrument it can be negative. We shall assume that $X_t$ obeys a time-homogeneous diffusion:
\begin{equation}
dX_t = \mu(X_t)\, dt + \sigma(X_t) \, dW_t.
\label{eq:assetdyn}
\end{equation}

Let $\theta_t$ be the allocation at time $t$ to $X$. Define the value function
\[
V_t =  \int_{s=t}^\infty  e^{-r(s-t)} \ex\big[\util(\theta_s  dX_s) \big]
\]
which\footnote{This is understood to be an It\^o integral, i.e.\ the increment $dX_s$ is `after' $\theta_s$, thus: $\theta_s(X_{s+ds}-X_s)$.} measures discounted expected utility of \emph{changes in} P\&L stretching forward to an infinite horizon. As $X_t$ follows a diffusion, we can simply perform the formal expansion
\[
\util(\theta_s  dX_s)  \equiv \util(0)+\theta_s  dX_s \, \util'(0) + \shalf \theta_s^2(dX_s)^2 \, \util''(0) + o(ds),
\]
so that we only care about the first and second derivatives of $\util$ at the origin. This is a departure from utility of wealth, in which the whole of the utility curve will be explored.
We stipulate $\util(0)=0$, $\util'(0)=1$, $\util''(0)=-1/\money$, so that $\money$, which is constant, is a measure of risk appetite and has units of money (because $\util$ does, in our formulation). It is natural to query why $\money$ should be constant. The reason is that, in general, a fund will operate many strategies, with each one allocated a risk budget: the portfolio manager will specify $\money$ for each strategy. Occasionally, $\money$ will need to be altered, depending on the amount of investment or redemption in the fund, on the fund's overall performance, and on the desired style balance. However, in between such re-gearing operations each strategy is to run a fixed level of risk, and this is our setup.

We have
\[
V_t = \ex_t\big[\util(\theta_t dX_t) \big] + (1-r\,dt) \ex_t[V_{t+dt}].
\]
Write $V_t=f(X_t)$, as the value function is not explicitly a function of calendar time, and let $dt\to0$. Expanding the expectation using It\^o's lemma (or the usual Feynman-Kac argument prevalent in option theory) gives 
%, to obtain a functional equation for $f$:
%\[
%f(X_t) = \dot{U}(X_t,\theta)\,dt  + (1-r\,dt) \left( f(X_t) + \mu(X_t) \pderiv{f}{x}\,dt + \shalf \sigma(X_t)^2 \pdderiv{f}{x} \,dt \right),
%\]
\begin{equation}
-r f(X_t) + \mu(X_t) \pderiv{f}{x} + \shalf \sigma(X_t)^2 \pdderiv{f}{x} = -\dot{U}\big(X_t,\theta(X_t)\big),
\label{eq:ode1}
\end{equation}
where
\[
\dot{U}(x,\theta) = \frac{1}{dt} \ex_t\big[\util(\theta  dX_t) \cdl X_t=x \big]
= \mu(x)\theta  - \frac{\sigma(x)^2\theta^2}{2\money}
\]
is the rate of accumulation of expected utility.
%We are to solve the ODE (\ref{eq:ode1}).

Write $\theta_t=g(X_t)$.
It is intuitively clear that the optimal allocation to the asset, in the absence of costs, is the value of $\theta$ that maximises the incremental utility $\dot{U}(x,g(x))$, i.e.\ 
\begin{equation}
\hat{\theta}_t \equiv \hat{g}_0(X_t) = \frac{\mu(X_t)}{\sigma(X_t)^2} \money.
\label{eq:theta_ntc}
\end{equation}
(The $_0$ in $\hat{g}_0$ indicates the transaction-free solution, and the $\hat{ }$ denotes optimality.) This is of the familiar form ``expected return $\div$ variance, $\times$ gearing factor''.
As $(\partial_2 \dot{U})(x,\hat{g}_0(x))=0$, we have that the `rebalancing gamma' (sensitivity of optimal no-cost position to change in price of instrument) is
\begin{equation}
\hat{g}_0'(X_t) = -\frac{(\partial_1\partial_2 \dot{U})(X_t,\hat{g}_0(X_t))}{(\partial_2^2\dot{U})(X_t,\hat{g}_0(X_t))} 
=  \frac{\mu'(X_t)-2\mu(X_t)\sigma'(X_t)/\sigma(X_t)}{\sigma(X_t)^2} \money ,
\label{eq:thetad_ntc}
\end{equation}
a result that we will need later as the width of the optimal NT zone is linked to it.
%The same result is obtained from solving for the integrated expected utility and then maximising.

We need some further notation. Write
\begin{equation}
\LL[f] \equiv \mu(x) \pderiv{f}{x} + \shalf \sigma(x)^2 \pdderiv{f}{x}
\label{eq:defL}
\end{equation}
for the infinitesimal generator of the diffusion. Then (\ref{eq:ode1}) becomes
\begin{equation}
(-r + \LL) [f](X_t)  =  - \dot{U}(X_t,g(X_t)).
\label{eq:basic}
\end{equation}
Solution of linear ODEs boils down to calculation of the complementary function (equation with RHS replaced with zero), which we denote $C_\pm(x)$, and the Green's function (equation with RHS with a delta-function at $x=\xi$, for any $\xi$), which we denote $K(x,\xi)$:
\begin{equation}
(-r+\LL) C_\pm = 0; \qquad(-r  + \LL) K(x,\xi) = -\delta(x-\xi).
\label{eq:CK}
\end{equation}
We stipulate (see Appendix) that $C_+$ is positive and monotone increasing, $C_-$ positive and monotone decreasing. The value function is then given as
\begin{equation}
f(x) = \int_{-\infty}^\infty \dot{U}(\xi,g(\xi)) K(x,\xi)\,d\xi.
\label{eq:valfncostfree}
\end{equation}
It is easily established that the Green's function is positive (see Appendix). Therefore, maximising the full integrated utility is indeed achieved by maximising the incremental utility as we did above through the choice  (\ref{eq:theta_ntc}).

\section*{Effect of costs: Main results}

\subsection*{ODE for value function}

We are going to assume that the optimal solution to the linear cost problem is of DT-NT-DT type. In the NT zone, no trading occurs so the value function $V_t=f(X_t,\theta_t)$ obeys almost the same equation as before, to wit
\[
(-r+\LL) f(x,\theta) = -\dot{U}(x,\theta),
\]
but note very carefully that $f$ is now a function of two variables (with the differential operator $\LL$ acting on the first one), and that $\theta$ is not adjusted as $x$ moves: in (\ref{eq:valfncostfree}), by contrast, it is variable.
The solution can be written down immediately as ``particular solution $+$ complementary function'', i.e.
\begin{equation}
f(x,\theta) = \int_{-\infty}^\infty \dot{U}(\xi,\theta) K(x,\xi) \,d\xi + \alpha_+(\theta) C_+(x) + \alpha_-(\theta) C_-(x),
\label{eq:valfnNT}
\end{equation}
where $\alpha_\pm(\theta)$ are weights to be determined; their values depend on the geometry of the NT boundary because the expression is valid only in the NT zone. 

In the DT zone the position is different. As instantaneous rebalancing is performed, the market does not have time to move, and the value function is obtained by deducting the cost of transacting towards the NT boundary. So:
\begin{equation}
f(x,\theta)  = f(x,\theta^\sharp)  -   
\left\{ \begin{array}{ll}\tcmsell  |\theta^\sharp-\theta|, & \theta^\sharp \le \theta \\ \tcmbuy  |\theta^\sharp-\theta|, & \theta^\sharp \ge \theta \end{array}\right. 
\label{eq:valfnDT}
\end{equation}
where $\theta^\sharp$, the target position, is the $\theta$-coordinate of the point on the nearer NT boundary vertically above or below the current position $(x,\theta)$ (as suggested by the arrows in Figure~\ref{fig:dtntdt}). The parameters $\tcmbuy $,$\tcmsell $, which have monetary units, are respectively the costs of buying and selling one lot of the asset, and we call them the TCMs (transaction cost multipliers).
 Importantly, therefore, the value function in the DT zones is determined immediately from its value on the boundaries.

We now need to glue the two solutions (\ref{eq:valfnNT},\ref{eq:valfnDT}) together. The value function is already continuous at the boundary by virtue of (\ref{eq:valfnDT}). Imposing differentiability in the $\theta$-direction (which we justify in the Appendix) gives
\begin{equation}
\left.
\begin{array}{rcl}
I(h_+(\theta),\theta) + \alpha'_+(\theta) C_+(h_+(\theta)) + \alpha'_-(\theta) C_-(h_+(\theta)) &=& -\tcmsell  \\
I(h_-(\theta),\theta) + \alpha'_+(\theta) C_+(h_-(\theta)) + \alpha'_-(\theta) C_-(h_-(\theta)) &=& \tcmbuy  
\end{array}
\right\}
\label{eq:bdcond}
\end{equation}
Here $h_+(\theta)$ is the value of $x$ satisfying $g_+(x)=\theta$, etc., and $I(x,\theta)$ is defined as
\begin{eqnarray*}
I(x,\theta) = \int_{-\infty}^\infty (\partial_2 \dot{U})(\xi,\theta) K(x,\xi)\, d\xi ,
% \\ I^*(x,\theta) &=& (\partial_1 I) (x,\theta),
\end{eqnarray*}
which obeys the ODE\footnote{Notation $\partial_2$ means differentiate (once) w.r.t\ the 2nd argument, and so on.}
\begin{equation}
(-r + \LL) I = -\partial_2 \dot{U}.
\label{eq:IDE}
\end{equation}

These matching conditions allow $\alpha_\pm(\theta)$ to be determined up to an arbitrary additive constant that can be identified from asymptotic behaviour.
Writing the determinant
\[
D(h_+,h_-) = \left| \begin{array}{cc}
C_+(h_+) & C_+(h_-) \\
C_-(h_+) & C_-(h_-)
 \end{array}\right| 
>0
\]
% note: D \approx (h_+-h_-) \times \mbox{Wronskian}
and noting that $\lim_{\theta\to+\infty} \alpha_-(\theta)=0$ and $\lim_{\theta\to-\infty} \alpha_+(\theta)=0$, we have
\begin{eqnarray}
\alpha_+(\theta) &=& \int_{-\infty}^\theta \frac{1}{D(h_+,h_-)} \Big[  -\tcmsell  C_-(h_-(\vartheta )) - \tcmbuy  C_-(h_+(\vartheta )) \nonumber\\
&& \mbox{}  + I(h_-(\vartheta),\vartheta) C_-(h_+(\vartheta )) - I(h_+(\vartheta),\vartheta) C_-(h_-(\vartheta )) \Big] \, d\vartheta ,\label{eq:obp0+}\\
\alpha_-(\theta) &=& \int_\theta^\infty \frac{1}{D(h_+,h_-)}\Big[- \tcmsell  C_+(h_-(\vartheta )) - \tcmbuy  C_+(h_+(\vartheta )) \nonumber\\
&& \mbox{} + I(h_-(\vartheta),\vartheta) C_+(h_+(\vartheta )) - I(h_+(\vartheta),\vartheta) C_+(h_-(\vartheta )) \Big]\, d\vartheta .\label{eq:obp0-}
\end{eqnarray}

Inserting these into (\ref{eq:valfnNT}) finalises the expression for the value function, which as far as we are aware is a new result. One sees immediately that if the NT zone width is shrunk to zero then $D(h_+,h_-)\to0$ and the integrand becomes singular ($-\infty$). This confirms that in the absence of a no-trade zone, infinitely much value is lost through frictional costs, as anticipated.

\subsection*{Optimal boundary equations}

We now have via (\ref{eq:valfnNT},\ref{eq:obp0+},\ref{eq:obp0-}) the value function for a given boundary, and the idea now is to optimise the boundary. To do this we need only to maximise the part of (\ref{eq:valfnNT}) that depends on the boundary location, i.e.\
\[
\alpha_+(\theta) C_+(x) + \alpha_-(\theta) C_-(x).
\]
The part of this that depends on $h_\pm(\theta)$ is $\alpha_\pm(\theta)$ which in turn are maximised by maximising the integrands of (\ref{eq:obp0+},\ref{eq:obp0-}).
Differentiating these gives the positions of the optimal boundaries (denoted $\hat{h}_\pm$):
\begin{eqnarray}
&& \pderiv{\alpha_+'}{h_+}= 0 \Leftrightarrow \pderiv{\alpha_-'}{h_+}= 0 \Leftrightarrow  \nonumber \\
&&\qquad \big[ C'_+(\hat{h}_+)\big(\tcmbuy  C_-(\hat{h}_+)+\tcmsell C_-(\hat{h}_-)\big)-C'_-(\hat{h}_+)\big(\tcmbuy C_+(\hat{h}_+)+ \tcmsell C_+(\hat{h}_-)\big)\big] \nonumber\\
&&\qquad \mbox{}  + I(\hat{h}_+,\theta) \big(C_+'(\hat{h}_+)C_-(\hat{h}_-)-C_-'(\hat{h}_+)C_+(\hat{h}_-)\big) \nonumber\\
&&\qquad \mbox{}  + I(\hat{h}_-,\theta) \big(C_-'(\hat{h}_+)C_+(\hat{h}_+)-C_+'(\hat{h}_+)C_-(\hat{h}_+)\big) \nonumber\\
&&\qquad \mbox{}  - (\partial_1I)(\hat{h}_+,\theta) D(\hat{h}_+,\hat{h}_-) =0 \label{eq:obp1+}
\end{eqnarray}
and
\begin{eqnarray}
&& \pderiv{\alpha_+'}{h_-}= 0 \Leftrightarrow \pderiv{\alpha_-'}{h_-}= 0 \Leftrightarrow  \nonumber \\
&&\qquad \big[ C'_-(\hat{h}_-)\big(\tcmbuy C_+(\hat{h}_+)+\tcmsell C_+(\hat{h}_-)\big)-C'_+(\hat{h}_-)\big(\tcmbuy C_-(\hat{h}_+)+\tcmsell  C_-(\hat{h}_-)\big)\big] \nonumber\\
&&\qquad \mbox{}  + I(\hat{h}_+,\theta) \big(C_-'(\hat{h}_-)C_+(\hat{h}_-)-C_+'(\hat{h}_-)C_-(\hat{h}_-)\big) \nonumber\\
&&\qquad \mbox{}  + I(\hat{h}_-,\theta) \big(C_+'(\hat{h}_-)C_-(\hat{h}_+)-C_-'(\hat{h}_-)C_+(\hat{h}_+)\big) \nonumber\\
&&\qquad \mbox{}  + (\partial_1I)(\hat{h}_-,\theta) D(\hat{h}_+,\hat{h}_-) =0 \label{eq:obp1-}
\end{eqnarray}
where for clarity we have abbreviated $h_\pm(\theta)$ to $h_\pm$.
These appear to be new results. They are a pair of coupled nonlinear (but not differential) equations which require the functions $C_\pm$, $C'_\pm$ and $I$ to be coded. For each $\theta$ their solution gives a pair $\big(h_-(\theta),h_+(\theta)\big)$ which demarcates the edges of the optimal NT boundary.

\subsection*{Small transaction costs}

In practice one would prefer not to have to solve the optimal boundary equations but rely on an approximation based on small $\varepsilon_\pm$. 
It turns out that the buying and selling costs only ever occur as their sum (intuitively: given long enough, the buys and sells cancel out, so the difference in TCM does not matter), so we write 
$
\varepsilon = \shalf(\tcmbuy+\tcmsell)
$.

Two sets of results can be obtained. The first comes from the difference $(\ref{eq:obp1+})-(\ref{eq:obp1-})$.
After much labour this reduces to an expression for the (horizontal) half-width of the NT zone,
\begin{equation}
\shalf\big(\hat{h}_+(\theta) - \hat{h}_-(\theta)\big) \sim \left(\frac{3\varepsilon\money}{2|\hat{g}_0'(x)|}\right)^{1/3} .
\label{eq:soln-gen}
\end{equation} 
The half-width in the vertical direction is obtained by multiplying by $|\hat{g}'_0(x)|$:
\begin{equation}
\shalf\big(\hat{g}_+(x) - \hat{g}_-(x)\big)  \sim \left(\frac{3\varepsilon \money |\hat{g}_0'(x)|^2}{2}\right)^{1/3}.
\label{eq:soln-gen2}
\end{equation} 
Two things may be seen immediately. The first is the one-third power law dependence on transaction cost. The second is the appearance of the `rebalancing gamma' $\hat{g}_0'(x)$ which we introduced previously and is given explicitly in terms of the drift and volatility by (\ref{eq:thetad_ntc}) without the need for solving any equations.
The cube root law agrees with Shreve \& Soner's derivation for the Merton problem\footnote{Stated in their Appendix,  \cite{Shreve94}.}. Incidentally the next terms in the expansions are $O(\varepsilon)$, $O(\varepsilon^{5/3})$ and so on.

The sum $(\ref{eq:obp1+})+(\ref{eq:obp1-})$ gives our second result, which informs of the horizontal displacement of the mid-point $\shalf\big(\hat{h}_+(\theta) + \hat{h}_-(\theta)\big)$ from its position in the absence of costs, $X=\hat{h}_0(\theta)$ (so $\hat{h}_0(\hat{g}_0(x))=x$).
This is found, after similar effort, to be:
\begin{equation}
\shalf\big(\hat{h}_+(\theta) + \hat{h}_-(\theta)\big) - \hat{h}_0(\theta)  \sim -\theta \left(\frac{2\varepsilon^2}{3|\hat{g}_0'(x)|^2\money}\right)^{1/3}, 
\label{eq:soln+gen}
\end{equation}
and the vertical displacement is
\begin{equation}
\shalf\big(\hat{g}_+(x) + \hat{g}_-(x)\big) - \hat{g}_0(x)  \sim -\theta \left(\frac{2\varepsilon^2|\hat{g}_0'(x)|}{3\money}\right)^{1/3};
\label{eq:soln+gen2}
\end{equation}
again, these can be calculated directly from (\ref{eq:thetad_ntc}).

Another question is, how much value is lost as a result of transaction costs? To answer this we have to look at the dependence of $f$ on $\varepsilon$.
Without optimising the NT boundary, the dependence of the value function on the transaction cost is simply $\propto \varepsilon$, as is obvious from (\ref{eq:obp0+},\ref{eq:obp0-}). More importantly, if we vary the transaction cost while simultaneously adjusting the NT boundary to keep it optimal, we see from (\ref{eq:obp0+},\ref{eq:obp0-}) that the numerator is $O(\varepsilon)$, as just pointed out, but the denominator is $O(\hat{h}_w)=O(\varepsilon^{1/3})$ on account of the $D(h_+,h_-)$ term. Hence the full dependence is $\propto\varepsilon^{2/3}$, which corroborates Shreve \& Soner's deductions \cite{Shreve94}.

\begin{figure}[h!]
\scalebox{0.85}{\includegraphics*{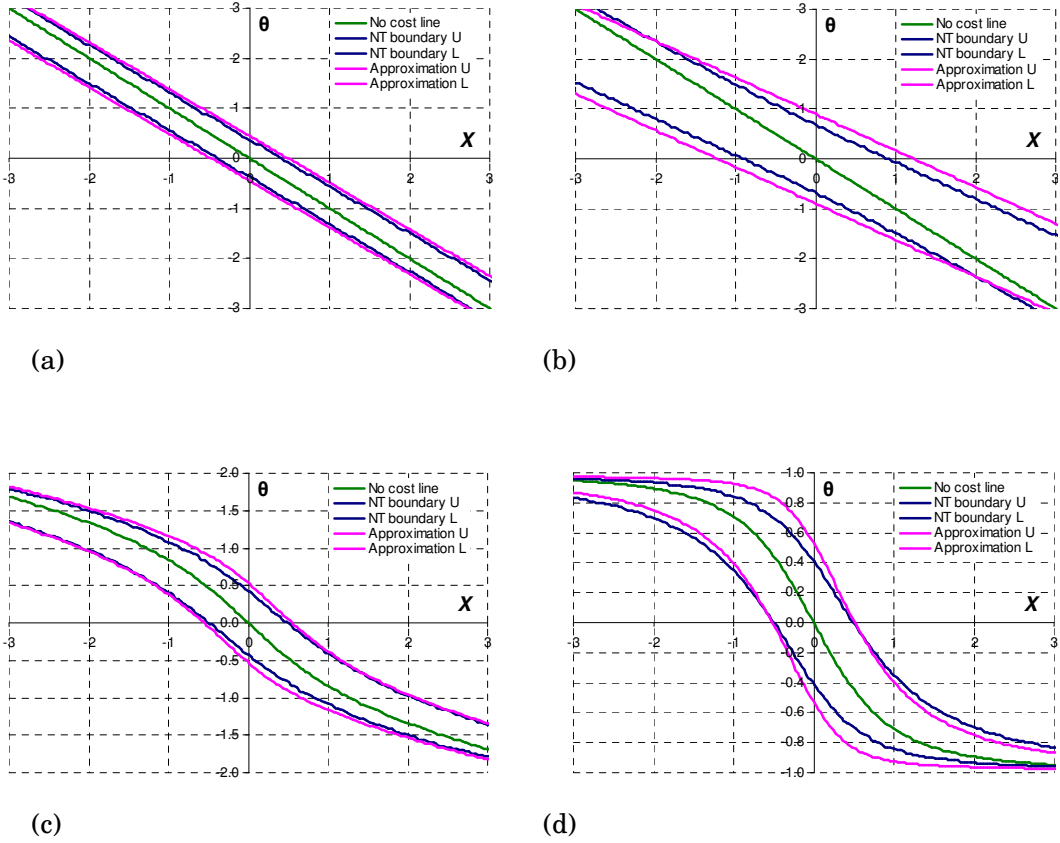}}
\caption{\small Boundaries of NT zone (dark blue lines) shown in relation to the transaction-free optimal allocation (green line) and perturbation approximation (pink lines). (a,b) are for OU process and (c,d) for extended OU process (\ref{eq:EOU}). Parameter values: (a-d) $\sigma=1$, $a=0$, $b=0.5$, $c=1$, $r=0.05$, $\money=2$. Local vol ($\nu$) values: (a,b) 0, (c) 0.125, (d) 0.25. TCM ($\varepsilon$) values: (a) 0.03, (b) 0.24, (c,d) 0.05.}
\label{fig:2}
\end{figure}

\section*{Numerical examples}

The Ornstein-Uhlenbeck process is the simplest model of a mean-reverting diffusive process and is specified by (see e.g.\ \cite{Oksendal92})
\[
dX_t = (a-bX_t) \, dt + \sigma\, dW_t, \qquad b>0.
\]
We can state our conclusions immediately. First $\hat{g}_0(x)=(a-bx)\money/\sigma^2$. Immediately we can find the half-widths of the NT zone in the horizontal and vertical directions from (\ref{eq:soln-gen},\ref{eq:soln-gen2}) as
\begin{equation}
\left(\frac{3\sigma^2\varepsilon}{2b}\right)^{1/3}, \qquad 
 \money \left(\frac{3b^2\varepsilon}{2\sigma^4}\right)^{1/3} 
\label{eq:soln-oueu}.
\end{equation} 
Using the expression for the displacement as well (\ref{eq:soln+gen2}), we obtain the approximate NT boundary as 
\[
\theta_\pm/G =\frac{a-bX}{\sigma^2}\left(1-
 \left(\frac{2b\varepsilon^2}{3\sigma^2}\right)^{1/3} \right)
 \pm  \left(\frac{3b^2\varepsilon}{2\sigma^4}\right)^{1/3} .
\]

For the purposes of demonstration we may as well assume $a=0$, $\sigma^2/2b=1$ so that the invariant measure of $X$ is standardised to $N(0,1)$, and that $\money=2$ so that $d\hat{\theta}/dX=-1$ in the transaction-free case ($\money$ only scales $\hat{\theta}$ so its effect is not interesting).  The only variables that matter are $\varepsilon$ and $r/b$ (in fact the latter does not even feature in the approximation, and it seems to have very little effect in practice).
To obtain the exact boundary we may either solve (\ref{eq:obp1+},\ref{eq:obp1-}), which requires $C_\pm$ to be computed (it is essentially a parabolic cylinder function in this case: see Appendix) or else use dynamic programming (also discussed in the Appendix). In the examples tested, the results were indistinguishable.

We graph the exact solution and this approximation in Figure~\ref{fig:2}(a,b) for two different cost levels: $\varepsilon=0.03$ (which in context is cheap) and 0.24 (expensive). As is apparent by eye from the figures, the optimal buffer width doubles as the TCM increases 8-fold. Also the agreement between the exact and approximated solutions is good.

The limit $b\to 0$ (no mean reversion) is interesting. The optimal costfree position is $aG/\sigma^2$ and the optimal strategy in the presence of transaction costs is to trade to the edge of the NT zone, specified by $\theta=(a\pm \varepsilon r)G/\sigma^2$ (this can be seen directly), \emph{and then hold the position without changing it}. Hence in (\ref{eq:soln-oueu}) the horizontal width $(3\sigma^2\varepsilon/2b)^{1/3}\to\infty$, and the vertical width $\money(3b^2\varepsilon/2\sigma^4)^{1/3}\to0$ because this is only the leading-order term: as there is no continuous rebalancing, a buffer of width $O(\varepsilon^{1/3})$ is unnecessary, so the next order term in the expansion, $O(\varepsilon)$, is the pertinent one.

A sort of `local volatility' extension to this model is given by
\begin{equation}
dX_t = -bX_t \, dt + \sigma \big(1+c^2X_t^2\big)^\nu \, dW_t.
\label{eq:EOU}
\end{equation}
The equilibrium point is fixed at zero and for $\nu>0$ the volatility increases away from it.
%In general there is little that can be done analytically, though with $\nu=\half$ the complementary functions are hypergeometric functions and the the invariant measure is (up to scaling) a Student-t with $\frac{2b}{c^2}+1$ d.f.
The no-cost line is no longer straight, as the increased volatility away from equilibrium makes larger positions unattractive. Figure~\ref{fig:2}(c,d) shows the results for this model, given by
\begin{eqnarray*}
\theta_\pm/G &=& \frac{-bX}{\sigma^2(1+c^2X^2)^{2\nu}}
+
\frac{bX}{\sigma^2} \left(\frac{2b\varepsilon^2}{3\sigma^2}\right)^{1/3} \frac{\big(1-4\nu c^2X^2(1+c^2X^2)^{-1}\big)^{1/3}}{(1+c^2X^2)^{8\nu/3}} 
\\
&& \pm  \left(\frac{3b^2\varepsilon}{2\sigma^4}\right)^{1/3} \frac{\big(1-4\nu c^2X^2(1+c^2X^2)^{-1}\big)^{2/3}}{(1+c^2X^2)^{4\nu/3}} ,
\end{eqnarray*}
 and again there is reasonable agreement between the perturbation expansion and the optimal boundary.
Notice that the buffer width, in the vertical sense, is thinner at the edges ($|X|$ large) than in the middle, as is seen from (\ref{eq:soln-gen2}).

\section*{Conclusions}

We have shown, in the case of a mean-reverting diffusion process, how to find the optimal solution of `DT-NT-DT' type and derived leading-order sensitivities to transaction cost. In common with the Merton problem we find that the optimal NT zone width is proportional to the cube root of the transaction cost. The results have been verified for the OU process and for an extension of it. An obvious extension is the multivariate case in which the dynamics of $X_t$ are driven by several extra factors; we are working on this.

An interesting phenomenon occurs when, in the extended-OU model (\ref{eq:EOU}), the parameter $\nu$ exceeds $\frac{1}{4}$. Then, the relationship between market level $X$ and optimal costfree position $\hat{g}_0(x)$ is no longer one-to-one, and so the relation $X=h(\theta)$, heavily exploited in the present analysis, is not well-defined: there is a particular problem at $X=\pm 1/(c\sqrt{4\nu-1})$, where $g'(X)=0$. Referring back to Figure~\ref{fig:dtntdt}, we have solved for the value function in the NT zone in horizontal slices $X\in[h_-(\theta),h_+(\theta)]$, but maybe a better approach would be to use vertical ones. However, we have not been able to do this, so it is a matter for further investigation.

\vspace{3mm}\noindent 

{\it Richard Martin and Torsten Sch\"oneborn are with AHL, part of Man Group PLC. The views expressed in this paper are their own rather than those of their institution. They thank Gunnar Klinkhammer and Thaleia Zariphopoulou for numerous helpful discussions and the referees for their suggestions for improvement. A longer version is available on request. Email {\tt rmartin@ahl.com}, {\tt tschoeneborn@ahl.com}}

\bibliographystyle{plain}
\bibliography{../phd}

%%%%%%%%%%%%%%%%%%%%%%%%%%%%%%%%%%%%%%%%%%%%%%%%%%%%%%%%%%%%%%%%%%%%%%%%%%%%%%%%%%%%%%%%%%%%%%%%%%%%%%%%%
%%%%%%%%%%%%%%%%%%%%%%%%%%%%%%%%%%%%%%%%%%%%%%%%%%%%%%%%%%%%%%%%%%%%%%%%%%%%%%%%%%%%%%%%%%%%%%%%%%%%%%%%%
\section*{Appendix}

\subsection*{Note on the complementary functions $C_\pm(x)$}

The existence of positive solutions to the equation $(-r+\LL)f=0$, one monotone increasing ($C_+$) and the other monotone decreasing ($C_-$), follows from standard theorems in ODE theory. Write $\psi(x)=f'(x)/f(x)$, a standard gambit, to obtain the Riccati equation
\[
-  \shalf \sigma(x)^2 \psi'(x) = \shalf \sigma(x)^2 \psi(x)^2 +\mu(x) \psi(x) - r.
\]
We claim that there exists a positive solution $\psi_+(x)$ and a negative solution $\psi_-(x)$ to this equation satisfying $\psi_+(-\infty)=\psi_-(+\infty)=0$.
This will prove the required statements about $C_\pm=e^{\int \psi_\pm}$.
 The RHS is a quadratic in $\psi$, and factorises as
\begin{equation}
\psi'(x) =  - \big( \psi(x) - \Psi_+(x) \big) \big( \psi(x)-\Psi_-(x) \big)
\label{eq:Riccati2}
\end{equation}
say, with $\Psi_+(x)>0>\Psi_-(x)$ (that the roots have opposite sign follows immediately from $r>0$). We only deal with $\psi_+$, as $\psi_-$ is analogous. By the Cauchy-Lipschitz theorem \cite{Ince28} there exists a unique positive solution to the Riccati equation satisfying $\psi_+(-\infty)=0$. If $0\le \psi_+(x) \le \Psi_+(x)$ then  $\mathrm{RHS}(\ref{eq:Riccati2})>0$ so $\psi_+(x)$ remains positive as $x$ increases; if $\psi_+(x)>\Psi_+(x)$ then it is still positive.

\subsection*{Note on the Green's function $K(x,\xi)$}

For $x<\xi$ and for $x>\xi$ the Green's function is just a solution to the homogeneous ODE, as the RHS of the differential equation $(-r+\LL)f=-\delta(x-\xi)$ is zero, but it is a different solution on each side\footnote{Currently the most accessible discussion is probably the Wikipedia entry for ``Green's function''.}. By integrating over a small segment at $x=\xi$, one finds that $K(x,\xi)$ is continuous in $x$ but its $x$-derivative jumps by $-2/ \sigma(\xi)^2$. Putting this together, we obtain
\[
K(x,\xi) =\left\{ \begin{array}{rr} C_-(\xi) C_+(x) , & x<\xi  \\
C_+(\xi) C_-(x) , & x>\xi  
\end{array} \right\} \times \frac{1}{\half\sigma(\xi)^2 \mathcal{W}\{C_-,C_+\}(\xi)} ,
\]
with $\mathcal{W}\{f,g\}\equiv fg'-gf'$ denoting the Wronskian. Positivity of $K$, as asserted previously, then follows immediately.

There is one issue that we have swept under the carpet: in the above construction we have implicitly dealt with the boundary conditions, by assuming that the Green's function decays at $\pm\infty$. The assumption is that the value function obeys the regularity conditions
\[
\lim_{x\to +\infty} f(x)/C_+(x) = 0; \qquad 
\lim_{x\to -\infty} f(x)/C_-(x) = 0.
\]
This is almost certainly true, but ought to be proven. Probably, it follows from a simple bounding argument on the value function.

\subsection*{Value function with costs; Dynamic programming}

Let $\theta_t^-$ denote the position \emph{before} rebalancing at time $t$. Then $\theta_t=\theta^-_{t+dt}$ and both are `known at time $t$'.
%as the rebalanced position $\theta_t$ is a deterministic function of $X_t$ and $\theta^-_t$. 
If we now have $V_t = f (X_t,\theta^-_t)$, then
\[
f(X_t,\theta^-_t) = (1-r\,dt) \ex_t[f(X_{t+dt},\theta_t)] + \dot{U}(X_t,\theta_t)\, dt - \left\{ \begin{array}{ll}\tcmsell  |\theta_t-\theta^-_t|, & \theta_t < \theta^-_t \\ \tcmbuy  |\theta_t-\theta^-_t|, & \theta_t > \theta^-_t \end{array}\right. .
\]
%Provided $|\theta_t-\theta^-_t|=O(dt)$, which seems plausible but might be wrong, we can expand the $\theta$-dependence as a differential. (Otherwise, we obtain an equation that is differential in $x$ and functional in $\theta$, which will be harder to handle.)
%This gives
%\footnote{It is understood that when acting on a function of two variables, $\LL$ acts on the first component only.}
%\[
%-r f(X_t,\theta^-_t)\, dt + \LL [f](X_t,\theta_t)\,dt   = -\dot{U}(X_t,\theta_t)\, dt  + \varepsilon |\theta_t-\theta^-_t|.
%\]
We take it as read that $\theta_t$ is a function $g$ of $X_t$ and $\theta_t^-$ only, i.e.\ a Markovian rebalancing strategy, and so we have
\begin{eqnarray}
f(x,\theta) &=& (1-r\, dt) \ex_t\big[ f\big(X_{t+dt},g(X_t,\theta)\big) \cdl X_t = x \big] \nonumber \\
&& \mbox{}+ \dot{U}(x,g\big(x,\theta)\big)\, dt  - 
\left\{ \begin{array}{ll}\tcmsell  |\theta_t-\theta^-_t|, & \theta_t < \theta^-_t \\ \tcmbuy  |\theta_t-\theta^-_t|, & \theta_t > \theta^-_t \end{array}\right. .
\label{eq:eqforvalfn}
\end{eqnarray}

A numerical method for solving (\ref{eq:eqforvalfn}) is dynamic programming. First, set up a grid $(X,\theta)$, and start with some approximation to $f$ such as $f\equiv 0$. Then perform the following iteration, which effectively is one time step of size $dt$, until convergence occurs (in $f$ and $g$): 
\begin{itemize}
\item
At each gridpoint $(X,\theta)$, proceed as follows:
\begin{itemize}
\item
Calculate for all possible rebalanced positions $g(X,\theta)$ on the grid the expectation term (approximately by observing that at a short horizon $X_{t+dt}$ is roughly Normally distributed with mean $\mu(X_t)$ and variance $\sigma(X_t)^2$). Also calculate the other two terms on the RHS of (\ref{eq:eqforvalfn}).
\item
Record which choice of $g(X,\theta)$ optimises the RHS of (\ref{eq:eqforvalfn}). 
\end{itemize}
\item
Repeat for all other gridpoints.
\end{itemize}
Convergence in $g$ occurs much more quickly than in $f$; indeed for zero transaction cost the convergence in $g$ occurs in one time step. The above iteration scheme defines a map $f\mapsto \mathfrak{T}[f]$ say, that improves the approximation of $f$ to the optimal value function. That convergence occurs in $f$ follows from the fact that $\mathfrak{T}$ is a contraction mapping, essentially because the interest rate is positive, so errors are slowly discounted. Notice that we have described the method for finding the value function and the optimal trading strategy, because we are maximising over $g$, but we do not have to do that: for example we might want to know about how much value is lost by deliberately using a specified (suboptimal) strategy. That calculation is faster of course as no search over $g$-values is required.

We recommend this method for finding $g$ (which is usually more important than $f$), mainly as a useful check on the analytical methods and approximations that we are about to derive.

\subsection*{\cutstart Note on differentiability at the boundary}

\begin{figure}[h!]
\centerline{\input{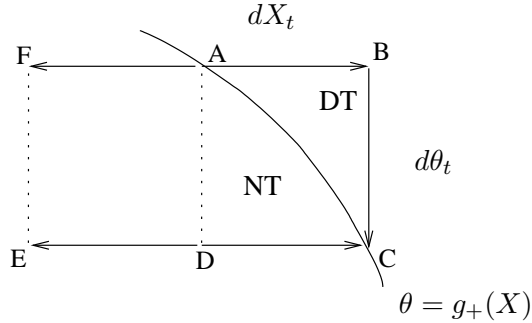}}
\caption{\small Sketch for `box argument' explaining why the value function is $\theta$-differentiable at the boundary.}
\label{fig:dtntdt_box}
\end{figure}

Referring to Figure \ref{fig:dtntdt_box}, which shows part of the upper NT-DT boundary, we wish to compare the value function at $A$ and at $D$, and claim that for an infinitesimal box $ABCD$, $V_A-V_D=\tcmsell  (\theta_D-\theta_A)$ to leading order. Consider the difference between being at $A$ and being at $D$, over the next time step $dt$. If $X$ falls in value ($A\rightarrow F$ or $D\rightarrow E$), no trading is done, so the only difference is through the P\&L which is slightly less for $A$ by an amount $|dX_t d\theta_t|=O(dt)$ (as one has a longer position at $A$ than at $D$). On the other hand if $X$ rises, in addition to the P\&L difference there is no transaction cost  from moving $D\rightarrow C$ as this is still in the NT zone, but there is a cost of $\tcmsell |d\theta_t|=O(\sqrt{dt})$ in moving $A\rightarrow B\rightarrow C$ as one has ended up in the DT zone. On the other hand, $V_A-V_D=(\partial f/\partial \theta)d\theta_t=O(\sqrt{dt})$, where in evaluating this one naturally uses the expression for $f$ inside the NT zone. Equating terms in $O(\sqrt{dt})$ gives $\partial f/\partial \theta=-\tcmsell $ i.e.\ (\ref{eq:bdcond}).
A similar argument holds for the lower boundary.

\subsection*{Note on the optimal boundary equations}

We refer to (\ref{eq:valfnNT}).
As the solution obtained by maximising
\[
\alpha_+(\theta) C_+(x) + \alpha_-(\theta) C_-(x)
\]
is to work for all $x$, we expect optimisation of $\alpha_+(\theta)$ and $\alpha_-(\theta)$ w.r.t.\ $\{h_-,h_+\}$ to give the same answer; furthermore, this answer should not depend on $\theta$ either. As $C_\pm(x)>0$, all we have to do is maximise $\alpha_\pm$.
That four equations reduce to two follows from the identities
\[
\frac{\partial \alpha'_+/\partial h_-}{\partial \alpha'_-/\partial h_-}
=-\frac{C_-(h_+)}{C_+(h_+)}, \qquad
\frac{\partial \alpha'_+/\partial h_+}{\partial \alpha'_-/\partial h_+}
=-\frac{C_-(h_-)}{C_+(h_-)}.
\]
It is necessary for such a reduction to occur, since otherwise we would have four equations for two unknowns, resulting in a solution that did not work for all values of $x$.
\cutend

\subsection*{Basic properties of the solution (for value function)}

As pointed out in the text, if $\varepsilon_\pm>0$, then $\alpha_\pm(\theta)\to-\infty$ and the value function tends to $-\infty$. 
So the optimal buffer width is certainly never zero. What is less clear is what happens for very large transaction costs.
There are conceivably two possibilities: (A) the value function increases sufficiently rapidly with market dislocation to absorb the transaction cost, and make it worth trading the asset when it is sufficiently far from equilibrium, even in very expensive markets; or (B) above some critical level of transaction cost, it is not worth trading. In the OU case, it is (A) that holds, and we suspect that this is generally true.

Next, we should be able to show the rather obvious result that in the absence of transaction costs, a buffer of positive width is suboptimal, but the suboptimality vanishes as the width is shrunk to zero. This is not as easy to prove as it may seem, at least by the methods we have derived here, so the following argument can possibly be improved.
The no transaction cost solution as previously derived is
\[
f_{NTC}(x) = \int_{-\infty}^\infty K(x,\xi) \dot{U}\big(\xi,\hat{g}_0(\xi)\big) \, d\xi.
\]
The solution we have derived for a buffered strategy is, on the other hand, on the line $\theta=\hat{g}_0(x)$, 
\[
f\big(x,\hat{g}_0(x)\big) = \int_{-\infty}^\infty K(x,\xi) \dot{U}\big(\xi,\hat{g}_0(x)\big) \, d\xi + \alpha_+\big(\hat{g}_0(x)\big) C_+(x) + \alpha_-\big(\hat{g}_0(x)\big) C_-(x) 
\]
(note very carefully that the integrand specifies $\hat{g}_0(x)$, whereas in $f_{NTC}$ it is $\hat{g}_0(\xi)$),
where $\alpha_\pm$ have now been found explicitly, and simplifiable given that $\varepsilon=0$. We are to show that $f\to f_{NTC}$, from \emph{below}, as the buffer width contracts to zero.
We put the buffer along the optimal NTC strategy and allow the width to contract to 0.

By the boundary conditions, of which we write \mbox{\bf [BC]} for the LHS, we have
\[
I\big(x,\hat{g}_0(x)\big) + \alpha'_+\big(\hat{g}_0(x)\big) C_+(x) + \alpha'_-\big(\hat{g}_0(x)\big) C_-(x) = 0 \quad \forall x,
\]
which can also be differentiated (we will use that presently, but not write it out here).

Notice first that
\[
(-r + \LL) f_{NTC}(x) = -\dot{U}\big(x,\hat{g}_0(x)\big)
\]
whereas
\begin{eqnarray*}
(-r + \LL_\bullet) f\big(x,\hat{g}_0(x)\big) &=& -\dot{U}\big(x,\hat{g}_0(x)\big)\\
&& \hspace{-3cm} \mbox{} + \alpha_+(\hat{g}_0(x))(-r+\LL)C_+(x) + \alpha_-(\hat{g}_0(x))(-r+\LL)C_-(x) \\
&& \hspace{-3cm} \mbox{} + \big[\mu(x)\hat{g}_0'(x) + \shalf \sigma(x)^2 \hat{g}_0''(x)\big] \mbox{\bf [BC]} \\
&& \hspace{-3cm} \mbox{} + \shalf \sigma^2(x)\hat{g}_0'(x) \frac{d}{dx}\mbox{\bf [BC]} \\
%&& \hspace{-3cm} \mbox{} + \shalf \sigma^2(x)\hat{g}_0'(x) \left[ \int_{-\infty}^\infty (\partial_1 K)(x,\xi)(\partial_2 \dot{U})(\xi,\hat{g}_0(x)) \, d\xi + \alpha_+'(\hat{g}_0(x))C_+'(x) + \alpha_-'(\hat{g}_0(x))C_-'(x)
&& \hspace{-3cm} \mbox{} + \shalf \sigma^2(x)\hat{g}_0'(x) \left[ (\partial_1 I)(x,\hat{g}_0(x)) + \alpha_+'(\hat{g}_0(x))C_+'(x) + \alpha_-'(\hat{g}_0(x))C_-'(x) \right]
\end{eqnarray*}
where the notation $\LL_\bullet$ emphasises that \emph{all} the $x$-dependence is being differentiated, i.e.\ that through the first argument of $f$ and through the second argument via $\hat{g}_0$. (This is what makes the algebra fairly messy, as twice differentiation of, for example, $\alpha(\hat{g}_0(x)) C(x)$ w.r.t.\ $x$ produces four terms.)
In the RHS of this expression, the second third and fourth lines are zero. To understand the last one, we take the limit  $h_\pm(\theta)\to \hat{h}_0(\theta)$, i.e. the NTC-optimal market value that corresponds to $\theta$, to obtain 
\begin{eqnarray*}
\alpha_+'(\theta) &=& \frac{-C_-(x)(\partial_1 I)(\hat{h}_0(\theta),\theta)+C_-'(\hat{h}_0(\theta))I(\hat{h}_0(\theta),\theta)}{\mathcal{W}\{C_+,C_-\}(\hat{h}_0(\theta))} \\
\alpha_-'(\theta) &=& \frac{C_+(x)(\partial_1 I)(\hat{h}_0(\theta),\theta)-C_+'(\hat{h}_0(\theta))I(\hat{h}_0(\theta),\theta)}{\mathcal{W}\{C_+,C_-\}(\hat{h}_0(\theta))} 
\end{eqnarray*}
But now when $\theta=\hat{g}_0(x)$, or equivalently $x=\hat{h}_0(\theta)$, we have
\[
\alpha_+'(\hat{g}_0(x)) C_+'(x) +  \alpha_-'(\hat{g}_0(x)) C_-'(x) = -(\partial_1 I)(x,\theta),
\]
so the fifth line also vanishes. Hence the two value functions $f$ and $f_{NTC}$, for a NT zone of zero width, and in the absence of transaction costs, differ by some element of $\mathrm{ker}[-r+\LL_\bullet]$, which we argue must be zero because of the behaviour at $x\to\pm\infty$.

An alternative route to this conclusion (which is messier, but more direct) is simply to express the Green's function in terms of the complementary functions, and hack out all the various integrals.

\subsection*{Limit of small transaction costs}

We study the solution for small $\varepsilon$ and in particular wish to study the width of the NT zone, as well as whether the NT zone is displaced relative to the costfree strategy $\theta=\hat{g}_0(X)$. By width we can either mean in the horizontal direction (`$X$-width') or vertical direction (`$\theta$-width').
We develop  (\ref{eq:obp1+},\ref{eq:obp1-}) around the point $(\tcmsell =\tcmbuy =0$, $h_+-h_-=0)$.
It is convenient to write
\[
h_w(\theta) = \shalf\big(h_+(\theta) - h_-(\theta)\big); \qquad h_m(\theta) = \shalf\big(h_+(\theta) + h_-(\theta)\big),
\]
with $w$ for half-\underline{w}idth and $m$ for \underline{m}id-point. Also, as the mid-point is near $\hat{h}_0$ (where $\hat{h}_0(\hat{g}_0(x))\equiv x$), we can split out the displacement of the mid-point from the transaction-free case as
\[
\hat{h}_m = \hat{h}_0+\hd,
\]
with $\hd$ ($d$ for \underline{d}isplacement) small.
%It is now fundamental to understand what orders all the various terms in (\ref{eq:obp1+},\ref{eq:obp1-}) are.
%We can do this from (\ref{eq:obp0+},\ref{eq:obp0-}) or (\ref{eq:obp1+},\ref{eq:obp1-}).

It is useful to recall some results from differential algebra. Define first
\[
\invar_{i,j} = C^{(i)}_+C^{(j)}_- - C^{(i)}_-C^{(j)}_+
\]
where superscripts $^{(i,j)}$ denote derivatives; this is a function of $x$ (and when necessary it will be evaluated at $x=\hat{h}_m$). Notice that
$
\invar_{1,0} = \mathcal{W}\{C_-,C_+\},
$ the Wronskian.
%where $\mathcal{W}$ denotes the Wronskian\footnote{$\mathcal{W}\{f,g\}\equiv fg'-gf'$} of the two solutions.
%about which we can say a couple of things: (i) it does not depend on $r$, being only a property of the diffusion, and (ii) it is the exponential-integral of the mean-to-variance ratio $-2\mu(x)/\sigma(x)^2$ (Abel's theorem; see Appendix for some more facts about generalised Wronskians).
%first, it is the reciprocal of the invariant measure, up to scaling %%% NO! %%%%, and is a property of the diffusion, not the interest rate; secondly, it is the and an important invariant of the differential equation (see Abel's theorem). 
An important result following directly from (\ref{eq:defL},\ref{eq:CK}) is
\begin{equation}
\frac{\invar_{2,1}}{\invar_{1,0}} = -\frac{r}{\half \sigma(x)^2}, \qquad \frac{\invar_{2,0}}{\invar_{1,0}} = -\frac{\mu(x)}{\half \sigma(x)^2}
\label{eq:Gamma012}
\end{equation}
(a principle which is expanded upon below) and so, in operator notation,
\[
\shalf \sigma(x)^2 \left( \invar_{2,1} - \invar_{2,0} \deriv{}{x} + \invar_{1,0} \dderiv{}{x}\right)
 \equiv \invar_{1,0} \cdot (-r + \LL).
\]
Furthermore, by differentiating again,
\[
\shalf \sigma(x)^2 \left( \invar_{3,1} - \invar_{3,0} \deriv{}{x} + \invar_{1,0} \ddderiv{}{x}\right) \equiv \invar_{1,0} \left(\deriv{}{x} - \frac{\mu(x)+\sigma'(x)\sigma(x)}{\half\sigma(x)^2} \right) (-r + \LL).
\]
This is important, as all the unwieldy expressions containing terms such as $C_+'''(x)C_-(x)-C_+(x)C_-'''(x)$, which arise in the Taylor series expansion of (\ref{eq:obp1+},\ref{eq:obp1-}), simplify to elementary functions of the coefficients of the underlying ODE, and hence to $\mu(x)$, $\sigma(x)$, $r$, and their derivatives, which are known immediately.

To make a start on the analysis we note first that
\[
D(\hat{h}_+,\hat{h}_-) = 2\hat{h}_w \invar_{1,0} + O(h_w^3)
\]
(by symmetry arguments there is no $O(h_w^2)$ term).
The next important point is that when we develop all the terms of (\ref{eq:obp1+},\ref{eq:obp1-}) in the vicinity of $\hat{h}_m$, the $O(h_w)$ term in the $I$-terms (i.e.\ the terms without $\varepsilon$ on the front) vanishes. Recall $\varepsilon := \shalf(\tcmsell +\tcmbuy )$.

\vspace{2mm}\noindent {\bf Difference $(\ref{eq:obp1+})-(\ref{eq:obp1-})$.} The symmetry of the non-$\varepsilon$ terms is odd, so we are interested in the cubic term. The expansion is
\[
4 \varepsilon \invar_{1,0} - {\textstyle  \frac{4}{3}} \big(  
\invar_{3,1} - \invar_{3,0} \partial_1 + \invar_{1,0} \partial_1^3 \big)
I(\hat{h}_m,\theta) \cdot \hat{h}_w^3 = O(\varepsilon\hat{h}_w^2,\hat{h}_w^5) 
\]
Recalling that $I(x,\theta)$ obeys (\ref{eq:IDE}), we can simplify the expression using this and (\ref{eq:Gamma012}) and the fact that $(\partial_2 \dot{U})(\hat{h}_0(\theta),\theta)=0$, to obtain
\[
4 \varepsilon \invar_{1,0} +   \frac{8(\partial_1\partial_2 \dot{U})(\hat{h}_0,\theta)}{3\sigma^2(\hat{h}_0)} \invar_{1,0}  \hat{h}_w^3 = O(\varepsilon \hat{h}_w^2, \hat{h}_w^5).
\]
%so
%\[
%\hat{h}_w \sim \left(\frac{3\sigma^2(\hat{h}_0)\varepsilon}{-2(\partial_1\partial_2 \dot{U})(\hat{h}_0(\theta),\theta)}\right)^{1/3} .
%\]
Using (\ref{eq:thetad_ntc}) we can simplify the second term on the LHS, and obtain (\ref{eq:soln-gen}).
%\begin{equation}
%\hat{h}_w \sim \left(\frac{3\varepsilon\money}{-2\hat{g}_0'(x)}\right)^{1/3} .
%\label{eq:soln-gen}
%\end{equation} 
%The half-$\theta$-width follows by multiplying by $|\hat{g}'_0(x)|$:
%\begin{equation}
%\hat{g}_w \sim \money \left(\frac{3\varepsilon}{2}\right)^{1/3} \left|\frac{\hat{g}_0'(x)}{\money}\right|^{2/3}.
%\label{eq:soln-gen2}
%\end{equation} 
%{\bf Therefore the $\theta$-width is proportional to the two-thirds power of the `rebalancing gamma' , and proportional to the cube root of the TCM.}
%Note that $\hat{g}$ is naturally proportional to $\money$ anyway, which is why we have grouped the $\money$ terms the way we have. 

\vspace{2mm}\noindent {\bf Sum $(\ref{eq:obp1+})+(\ref{eq:obp1-})$.} The symmetry of the non-$\varepsilon$ terms is even, and so we have
\[
4 \varepsilon \hat{h}_w \invar_{2,0} -4 \big(
\invar_{2,1} - \invar_{2,0} \partial_1 + \invar_{1,0} \partial_1^2  \big)
I(\hat{h}_m,\theta) \cdot \hat{h}_w^2
=  O(\varepsilon\hat{h}_w^2,\hat{h}_w^4) .
\]
The second term on the LHS emerges (again applying (\ref{eq:Gamma012},\ref{eq:IDE})) as
\[
\frac{4\hat{h}_w^2}{\half \sigma(\hat{h}_m)^2} \invar_{1,0} (\partial_2 \dot{U})(\hat{h}_m,\theta),
\]
so we obtain
\[
\mu(\hat{h}_m)\varepsilon  \sim (\partial_2 \dot{U})(\hat{h}_m,\theta) \, \hat{h}_w .
\]
Again we use $(\partial_2 \dot{U})(\hat{h}_0(\theta),\theta)=0$ to obtain 
\[
\mu(\hat{h}_0)\varepsilon  \sim  (\partial_1\partial_2 \dot{U})(\hat{h}_0,\theta) \hat{h}_w \hd,
\]
and we end up with (\ref{eq:soln+gen}).
%\begin{equation}
%\hd \sim -\frac{2\theta}{3\money} \hat{h}_w^2.
%\label{eq:soln+gen}
%\end{equation}
%{\bf Therefore the horizontal deviation of the median of the optimal NT zone, from the no-cost line, is proportional to the two-thirds power of the TCM.} 

\subsection*{\cutstart Notes on invariants of ODEs}

If
\[
\dderiv{y}{x} + p(x)\deriv{y}{x} + q(x) y = 0
\]
then, defining
\[
\invar_{i,j} \equiv y_1^{(i)}y_2^{(j)} - y_2^{(i)}y_1^{(j)},
\]
we have
\[
\frac{\invar_{2,0}}{\invar_{1,0}} = -p; \qquad
\frac{\invar_{2,1}}{\invar_{1,0}} = q; \qquad
\frac{\invar_{3,0}}{\invar_{1,0}} = p^2-p'-q ;\qquad
\frac{\invar_{3,1}}{\invar_{1,0}} = q'-pq.
\]
Note that $\invar_{1,0}=-\mathcal{W}\{y_1,y_2\}$ with $\mathcal W$ denoting the Wronskian.
The first two results are somewhat analogous to the sum and product formulas for the roots of a quadratic. 
The analogy is made obvious when one considers $y_i(x)=e^{\lambda_i x}$, $i=1,2$, in which case $-p$ and $q$ are constants and equal to the sum and the product of the $(\lambda_i).$

More generally, one can recognise that $\invar_{i,j}/\invar_{k,l}$ is always a rational function of the coefficients of the ODE ($p$, $q$) and their derivatives. This is because it is invariant under any transformation of the form $y_1\mapsto ay_1+by_2$, $y_2\mapsto cy_1+dy_2$ with $ad-bc\ne0$, i.e.\ the action of the matrix group $\mathbf{GL}_2(\C)$. Hence it is also invariant under the action of the differential Galois group\footnote{See e.g.\ I. Kaplansky, \emph{An Introduction to Differential Algebra}, Hermann, Paris, 1957.} of the differential field extension $\C\langle x,p,q,y_1,y_2\rangle/\C\langle x,p,q\rangle$ (this group being a subgroup of $\mathbf{GL}_2(\C)$), and therefore must be contained in $\C\langle x,p,q\rangle$. In the same way, in the case of a quadratic equation $y^2+py+q=0$, with roots $y_{1,2}$, expressions that are symmetric under $y_1\leftrightarrow y_2$, such as $\frac{y_1^3+y_2^3}{y_1^2+y_2^2}$, are expressible as rational functions of $p$ and $q$, whereas nonsymmetric ones, such as $y_1+2y_2$, are not.
\cutend

\subsection*{Notes on the OU case and on $\mathbf{D}^\pm_\nu$}

Some more details of the OU case are now stated for completeness.
The ODE (\ref{eq:basic}) is easily solved to give the costfree value function as
\begin{equation}
\hat{f}_0(x) =\frac{\money b}{2(r+2b)} \!\left(\shalf \mathfrak{z}^2 + \frac{b}{r}\right), \qquad \mathfrak{z} \equiv \frac{x-a/b}{\sigma/\sqrt{2b}}.
\end{equation}
Note that $\mathfrak{z}$ is the `z-score', and $b/r$ is the price of a riskfree perpetual bond paying a coupon of $b$.
%The whole lot boils down to an expression in the nondimensional quantities $b/r$ (mean reversion rate normalised by discount rate) and the z-score, all multiplied by the scaling factor $\money$.
%When there is no mean reversion, $b=0$ and we get the value function as $\hat{f}_0(x) = a^2\money/2\sigma^2 r$ independently of $x$.
%(See note\footnote{NT zone then has horizontal boundaries at $\theta=(a\pm r\varepsilon) \money/\sigma^2$.}.)
Also $I(x,\theta)$ is given by
\[
I(x,\theta) = - \frac{x-a/b}{1+r/b} - \frac{\sigma^2 \theta}{r\money} ; 
\qquad (\partial_1I)(x,\theta)=\frac{-1}{1+r/b}.
\]
The complementary functions $C_\pm(x)$ are given by
\begin{equation}
C_\pm(x) = \mathbf{D}^\pm_{r/b}\left(\frac{x-a/b}{\sigma/\sqrt{2b}}\right), \qquad 
\mathbf{D}_\nu^\pm (x) := \int_0^\infty e^{\pm zx} z^{\nu-1} e^{-z^2/2} \, dz.
\label{eq:Dnu}
\end{equation}
The functions $\mathbf{D}_\nu^\pm$ are related to the parabolic cylinder functions, and here are some of their properties, by which the numerical implementation of the integral can be checked:
\begin{itemize}
\item
$\mathbf{D}^\pm_\nu(0)=2^{\nu/2-1}\Gamma(\frac{\nu}{2})$
\item
${\mathbf{D}^\pm_\nu}'(0)=\pm 2^{(\nu-1)/2}\Gamma(\frac{\nu+1}{2})$
\item
$\mathbf{D}^+_\nu(x)\sim \Gamma(\nu)/(-x)^\nu$ for $x\to-\infty$  (symmetrically for $\mathbf{D}^-_\nu$).
\item
$\mathbf{D}^+_\nu(x)\sim \sqrt{2\pi}x^{\nu-1}e^{x^2/2}$ for $x\to+\infty$  (symmetrically for $\mathbf{D}^-_\nu$).
\item
For small $\nu>0$ we have $\mathbf{D}^\pm_\nu(x) \approx \frac{1}{\nu} + \int_0^\infty (\ln z )(z-x)e^{\pm zx-z^2/2} \, dz$.
\end{itemize}
(NB: Our definition (\ref{eq:Dnu}) is convenient, but nonstandard; see also \cite{Abramowitz64}.)
The Wronskian is, with the chosen normalisation,
\[
\mathcal{W}\{C_-,C_+\}(x) = \Gamma(r/b) \big/ \phi(\mathfrak{z}), 
\]
with $\mathfrak{z}$ denoting the z-score as above, and $\phi$ the standard Normal pdf.

\end{document}